\documentstyle{article}
\font\cero=cmss10 scaled 1728 \font\uno=cmssbx10 scaled 1200
\begin{document}
\begin{flushleft}
{\cero Wilson loops and topological phases in closed string
theory}
\\[3em]
\end{flushleft}
{\sf R. Cartas-Fuentevilla}\\
{\it Instituto de F\'{\i}sica, Universidad Aut\'onoma de Puebla,
Apartado postal J-48 72570, Puebla Pue., M\'exico
(rcartas@sirio.ifuap.buap.mx).}  \\

Using covariant phase space formulations for the natural
topological invariants associated with the world-surface in closed
string theory, we find that certain Wilson loops defined on the
world-surface and that preserve topological invariance, correspond
to wave functionals for the vacuum state with zero energy. The
differences and similarities with the 2-dimensional QED proposed
by Schwinger early are discussed. \\

\noindent {\uno I. Introduction} \vspace{1em}

At present one of the fundamental problems in string theory is the
understanding of its non-perturbative aspects, intimately related
with the search for the underlying geometry of the theory, which
will allow us, for example, to determine the corresponding true
vacuum. In this sense, the developed of topological field theory
constitutes an important step in such a direction, specifically in
the construction of the called {\it unbroken phase}, where the
relevant symmetries of the theory such as general covariance are
unbroken \cite{1}.

Although it is undeniable the fundamental role that the
topological invariants of the theory play in this context, it is
strange that we have not yet a more profound understanding of the
underlying geometry of those topological invariants that emerge in
a natural way in string theory, i.e. the Euler characteristic and
the second Chern number of the worldsheet. However in \cite{2} we
have demonstrated that the underlying geometrical structure in a
covariant phase space formulation for such topological string
actions mimics that of a two-dimensional U(1)-gauge theory, whose
geometrical meaning is perhaps the most understood. Therefore, in
the present work we attempt to exploit the results in \cite{2} and
to gain insight into the geometrical structure of these
topological invariants and their intimate relations with the
vacuum state of the theory.

The main result will be that in a second quantization scheme for
these topological string actions, the wave functionals for the
vacuum state with zero energy correspond in a natural  way to the
Wilson loops of certain connections characterizing the
two-dimensional world-surface along the spatial configuration of
the closed string.

This work is organized as follows. In the next section we give an
outline on the two-dimensional world-surface that will be useful
in the developed of the work. In Section III we describe briefly
the covariant phase space formulation given in \cite{2} for the
Euler characteristic. In Section IV the quantization and the
Wilson loops as wave functionals for the ground state  are
considered. In Section V the constraints and symmetries are
discussed, and subsequently in Section VI the covariance and
topological invariance of the Wilson loops are analyzed. The
second Chern number is considered in Section VII, and we finish in
Section VIII with some concluding remarks. \\

\noindent {\uno II. Preliminaries of the embedding 2-surface}
\vspace{1em}

It is convenient to give a survey on some basic ideas on the
embedding two-surface that will be useful in the present
treatment.

Considering the expression for the internal curvature tensor
$R_{\mu\nu\alpha\beta}$ of an imbedding given by Carter \cite{3}
in terms of an internal gauge connection $\rho$, we can find that,
for a two-dimensional embedding
\begin{equation}
     R_{\mu\nu} = R_{\mu\sigma\nu} {^{\sigma}} = \frac{1}{2}
     n_{\nu\alpha} {\cal E}^{\alpha\lambda} (\overline{\nabla}_{\mu}
     \rho_{\lambda} - n^{\pi}_{\mu} \overline{\nabla}_{\lambda}
     \rho_{\pi}),
\end{equation}
where $n^{\mu\nu}$ is the fundamental tensor of the imbedding
two-surface, which is characterized by the antisymmetric unit
tangent element tensor
\begin{equation}
     {\cal E}^{\mu\nu} = \iota^{\mu} \vartheta^{\nu} - \iota^{\nu}
     \vartheta^{\mu},
\end{equation}
$\iota^{\mu}$ being a timelike unit vector and $\vartheta^{\mu}$ a
spacelike unit vector, both tangential to the world-surface.
Furthermore, $\rho_{\sigma} = {\cal E}^{\nu} {_{\mu}} \
\rho_{\sigma} {^{\mu}}_{\nu}, $ where $\rho_{\sigma}
{^{\mu}}_{\nu}$ corresponds to the background spacetime components
of the internal frame components of the natural gauge connection
for the group of 2-dimensional internal frame rotations.
Furthermore, $\overline{\nabla}_{\mu} \equiv n^{\alpha}_{\mu}
\nabla_{\alpha}$, where $\nabla_{\alpha}$ is the usual Riemannian
covariant differentiation operator associated with the background
metric $g_{\mu\nu}$.

From Eq.\ (1), we find that
\begin{equation}
     R_{00} = - \frac{1}{2} {\cal E}^{0i} (\overline{\nabla}_{0}
     \rho_{i} - \overline{\nabla}_{i} \rho_{0}),
\end{equation}
where we have considered that $n_{00} = 1$, $n_{0i} = 0$.
Additionally, in \cite{3} it is shown that for a 2-dimensional
world-surface
\begin{equation}
     R = \xi^{\mu\nu} \overline{\nabla}_{\mu} \rho_{\nu} = {\cal E}^{0i}
     (\overline{\nabla}_{0} \rho_{i} - \overline{\nabla}_{i}
     \rho_{0}) = \iota^{0} \vartheta^{i} (\overline{\nabla}_{0}
     \rho_{i} - \overline{\nabla}_{i} \rho_{0}),
\end{equation}
where Eq.\ (2) has been used. Equations (3), and (4) will be
useful below in the construction of the second quantization for
the Euler characteristic $\chi$ of the two-dimensional
world-surface $S$ embedded in an arbitrary background spacetime,
which can be described in terms of the inner curvature scalar $R$
as \cite{3}
\begin{equation}
     \chi= (2-2g) = \sigma_{1} \int_{S} \sqrt{-\gamma} R dS,
\end{equation}
where $g$ is the number of handles of the world surface, and
$\gamma$ the determinant of the embedded surface metric. In
\cite{2} it is shown that
\begin{equation}
     \delta \ \chi = \sigma_{1} \int_{S} \sqrt{-\gamma} \ \left(
     \frac{1}{2} R \ n^{\mu\nu} - R^{\mu\nu} \right) \delta
     g_{\mu\nu} \ dS,
\end{equation}
modulo a total divergence; hence the energy-momentum tensor
vanishes
\begin{equation}
     T^{\mu\nu} = \frac{1}{2} R n^{\mu\nu} - R^{\mu\nu} = 0,
\end{equation}
as expected for a topological invariant, since
\begin{equation}
     \frac{1}{2} R n^{\mu\nu} - R^{\mu\nu} = 0,
\end{equation}
identically for a two-dimensional surface \cite{3}. Although it is
evident that at classical level a topological invariant is
physically trivial, we shall see that at quantum level the things
can be very different.  \\

\noindent {\uno III. The phase space formulation for the Euler
characteristic}
\vspace{1em}

In \cite{2} it is shown that the nontrivial phase space
formulation for $\chi$ is given by a covariant and gauge-invariant
symplectic structure
\begin{equation}
     \omega = \int_{\Sigma} \delta (\sqrt{-\gamma}
     {\cal E}^{\mu\nu}) \delta \rho_{\nu} d
     \overline{\Sigma}_{\mu},
\end{equation}
the constraint
\begin{equation}
     \overline{\nabla}_{\mu} {\cal E}^{\mu\nu} = 0,
\end{equation}
and the {\it Bianchi identity}
\begin{equation}
     R \ {\cal E}_{\kappa\lambda} = 2 n_{[\lambda}{^{\sigma}}
     \overline{\nabla}_{\kappa ]} \ \rho_{\sigma},
\end{equation}
which gives the two-form $R {\cal E}$ as the {\it exterior
derivative} of the one-form $\rho$ \cite{3}.

Equations (9)-(11) mimic in its mathematical structure and
symmetry properties the phase space formulation of an Abelian
gauge theory \cite{2}. However, there exists an important
difference since, unlike the conventional (3+1)-dimensional
$U(1)$-gauge theory, Eq.\ (9)-(11) represents a (1+1)-dimensional
$U(1)$-gauge theory {\it embedded} in an arbitrary ambient
spacetime. In relation to the dimensionality, Eqs.\ (9)-(11) would
be closer to the (1+1)-dimensional QED (without massless fermions)
proposed by Schwinger early \cite{4}, and whose first Hamiltonian
analysis and solutions were given for example in \cite{5}. The
present analysis of the canonical formulation for $\chi$ will have
some parallelism with \cite{5} although,  of course, also its
particularities.

For convenience we shall make a decomposition space+time of the
covariant phase space formulation of $\chi$ given by Eqs.\
(9)-(11). The covariance will be recuperated at the end.

The choice of $\iota^{\mu}$ as a timelike vector, and
$\vartheta^{\mu}$ as a spacelike one, induces naturally a
(1+1)-decomposition on the covariant description of the
formulation (9)-(11); such a decomposition was already used in the
expressions (3),  and (4).

In this manner, considering that in Eq.\ (9)
$d\overline{\Sigma}_{\mu}$ is normal to the Cauchy spacelike
surface $\Sigma$, we can choose $d\overline{\Sigma}_{\mu} =
\iota_{\mu} d\Sigma$, and $\omega$ takes the noncovariant form
\begin{equation}
     \omega = \int_{\Sigma} \delta (\sqrt{-\gamma}
     {\cal E}^{0i}) \delta \rho_{i} d \Sigma,
\end{equation}
which allows us to use a {\it temporal gauge}
\begin{equation}
     \rho_{0} = 0,
\end{equation}
in order to simplify our calculations. Although $\Sigma$ is
strictly a Cauchy hyper-surface, the integral on $\Sigma$ can be
reduced actually to an integral on the 1-dimensional spatial
configuration of the closed string, as we shall do below.

Finally, Eq.\ (10) implies that
\begin{equation}
     \overline{\nabla}_{i} {\cal E}^{i0} = 0,
\end{equation}
in this noncovariant description of the phase space. Equation (14)
would represent ``the Gauss law constraint", the analogue of that
of the conventional gauge theory. \\

\noindent {\uno IV. Second quantization and Wilson loops}
\vspace{1em}

Equation (12) shows explicitly that the canonically conjugate
phase space variables are $(\sqrt{-\gamma} \iota^{0}) {\cal
E}^{0i}$, and the spatial connection $\rho_{i}$. In this manner,
in order to make the corresponding quantization we have the
correspondence
\begin{eqnarray}
     \rho_{i} \!\! & \rightarrow & \!\! \rho_{i}, \nonumber \\
     {\cal E}^{0i} \!\! & \rightarrow & \!\! (\iota^{0}) i
     \frac{\delta}{\delta\rho_{i}}.
\end{eqnarray}
For constructing the quantum Hamiltonian $(H_{Q})$ we can use the
classical expression for $T^{00}$ given by Eqs.\ (3), (4), and
(8),
\begin{equation}
     T^{00} = \frac{1}{2} [ {\cal E}^{0i} \overline{\nabla}_{0}
     \rho_{i} - \iota^{0} \vartheta^{i} \overline{\nabla}_{0}
     \rho_{i}],
\end{equation}
and then,
\begin{equation}
     H_{Q} = \int_{\Sigma} F_{i} \big( i
     \frac{\delta}{\delta\rho_{i}} - \vartheta_{i} \big) d\Sigma,
\end{equation}
where $F_{i} = \iota^{0} \overline{\nabla}_{0} \rho_{i}$; since
$[F_{i}, \frac{\delta}{\delta\rho_{i}}] = 0$, we have not ordering
ambiguity in the Hamiltonian (17).

Furthermore, the ground state wave functionals $\psi$, a
representation of the vacuum state of the theory, satisfy $H_{Q}
\psi = E \psi$, and then for a state of zero energy $\psi$
satisfies
\begin{equation}
     \big( i \frac{\delta}{\delta\rho_{i}} - \vartheta_{i} \big) \psi
     = 0,
\end{equation}
for which the unique solution will be given evidently by
\begin{equation}
     \psi(\rho) = Ae^{-i \int_{\Sigma} \vartheta_{i} \rho_{i}
     d\Sigma},
\end{equation}
where $A$ is a constant parameter. Considering that
$\vartheta_{i}$ is a space-like vector tangent to the
world-surface, then $\vartheta_{i}$ goes tangentially along the
closed string loop, and hence
\begin{equation}
     \psi(\rho) = Ae^{-i \oint \rho_{i} d\vartheta_{i}}.
\end{equation}
In this manner, our ground state wave-function corresponds, in a
natural way, to the Wilson loop for the connection $\rho_{i}$.
Note that $\rho_{i}$ is of support confined just on the closed
string loop. \\

\noindent {\uno V. Constraints and symmetries}
\vspace{1em}

In \cite{2} it is proved that the action (5), and its covariant
phase space formulation given in Eqs.\ (9)-(11) are invariant
under the {\it gauge transformation} of the connection
$\rho_{\mu}$,
\begin{equation}
     \rho_{\nu} \rightarrow \rho_{\nu} + \overline{\nabla}_{\nu}
     \phi,
\end{equation}
where $\phi$ is an arbitrary scalar field. It is in this sense
that $\chi$ and its covariant canonical formulation mimic the
symmetry properties of a $U(1)$-gauge theory \cite{2}.

It is easy to see that under the {\it spatial restriction} of (21)
\begin{equation}
     \rho_{i} \rightarrow \rho_{i} + \overline{\nabla}_{i}
     \phi,
\end{equation}
the wave-function $\psi$ (20) is gauge invariant because the
string loop $\vartheta$ is closed. In this sense $\psi$ represents
a physical state for the theory.

Furthermore, the ``Gauss law" (14) must be imposed as a constraint
on the quantum state $\psi$,
\begin{equation}
     \overline{\nabla}_{i} \frac{\delta}{\delta\rho_{i}} \psi = 0,
\end{equation}
which implies that $\psi$ must be gauge invariant in the sense of
(22), such as in the conventional quantum gauge theory; thus, our
wave-function $\psi$ solves automatically the constraints of the
theory. In this manner, the ``Gauss law" (14) is the generator of
the gauge symmetry at quantum level.

At classical level ${\cal E}^{0i}$ is essentially $\vartheta^{i}$,
and in the present canonical analysis ${\cal E}^{0i}$ is the
variable conjugate to $\rho_{i}$,  and hence we have the primary
constraints on the physical states
\begin{equation}
      \big( i \frac{\delta}{\delta\rho_{i}} - \vartheta_{i} \big)
      \psi = 0,
\end{equation}
which corresponds exactly to the equation (18) for the
wave-function (20); thus $\psi(\rho)$ solves naturally the primary
constraints of the theory. In this sense, the quantum Hamiltonian
(17) is a pure combination of the primary constraints. \\

\noindent {\uno VI. Covariance and topological invariance of
$\psi(\rho)$}
\vspace{1em}

The decomposition space+time used previously is not strictly
necessary, and we have employ it only for computational
convenience and for making contact with the conventional non
covariant description of quantum gauge theory. However, we can get
back the covariance directly on the Wilson loops given in (19). It
is evident that $\psi(\rho)$ can be written as
\begin{equation}
     \psi(\rho) = Ae^{-i \int_{\Sigma} {\cal E}^{\mu\nu} \rho_{\mu}
     d\overline{\Sigma}_{\nu}},
\end{equation}
which is manifestly covariant. Note that now (25) is invariant
under the gauge transformation (21) since, considering Eq.\ (10)
\begin{equation}
     \int_{\Sigma} {\cal E}^{\mu\nu} (\rho_{\mu} +
     \overline{\nabla}_{\mu} \phi) d \Sigma_{\nu} = \int_{\Sigma}
     {\cal E}^{\mu\nu} \rho_{\mu} d\Sigma_{\nu} + \int_{\Sigma}
     \overline{\nabla}_{\nu} ({\cal E}^{\mu\nu} \phi)
     d\Sigma_{\nu},
\end{equation}
where the last integral can be reduced to
\begin{equation}
     \int_{\Sigma} \overline{\nabla}_{\nu} ({\cal E}^{\mu\nu}
     \phi) d\Sigma_{\nu} = \int_{\partial\Sigma} {\cal E}^{\mu\nu}
     \phi \ d\Sigma_{\mu\nu},
\end{equation}
which vanishes (on $\partial\Sigma$) for fields with compact
support in the {\it spatial directions}.

Furthermore, the covariant expression (25) shows manifestly that
the Wilson loop $\psi(\rho)$ does not depend on any background
metric structure, and is topological in character. In this manner,
the Wilson loop $\psi(\rho)$ describes a topological phase with
unbroken diffeomorphism invariance in string theory. \\

\noindent {\uno VII. Another topological invariant: the second
Chern number}
\vspace{1em}

For a string in ordinary 4-dimensional spacetime we have another
topological invariant for the world-surface, the second Chern
number of the normal bundle \cite{3}
\begin{equation}
     \chi' = \sigma_{2} \int_{S} \sqrt{-\gamma} \ \Omega \ d
     {S},
\end{equation}
where $S$ is the entire imbedding two-surface and $\sigma_{2}$ a
parameter; furthermore,
\begin{eqnarray}
     \Omega \!\! & = & \!\! \overline{\nabla}_{\mu} ({\cal
     E}^{\mu\nu} {\omega}_{\nu}), \nonumber \\
     \omega_{\nu} \!\! & = & \!\! \frac{1}{2} \omega_{\nu}
     {^{\mu\lambda}} \ \epsilon_{\lambda\mu\rho\sigma} \ {\cal
     E}^{\rho\sigma}, \nonumber
\end{eqnarray}
where $\omega_{\nu} {^{\mu\lambda}}$ corresponds to the external
frame rotation (pseudo-)tensor, and
$\epsilon_{\lambda\mu\rho\sigma}$ the antisymmetric background
measure tensor. $\omega_{\nu}$ is the outer analogue of
$\rho_{\mu}$, the inner gauge connection considered above.
Geometrically $\chi'$ is related with the number of
self-intersections of the two-surface.

In this case the covariant phase space formulation is given by the
symplectic structure
\begin{equation}
     \omega' = \int_{\Sigma} \delta (\sqrt{-\gamma}
     {\cal E}^{\mu\nu}) \delta \omega_{\nu} \ d
     \overline{\Sigma}_{\mu},
\end{equation}
the constraint (10), and the Bianchi identity
\begin{equation}
     \Omega {\cal E}_{\mu\nu} = 2 n_{[\nu} {^{\sigma}} \
     \overline{\nabla}_{\mu ]} \ \omega_{\sigma}.
\end{equation}

The corresponding {\it gauge transformation} of the connection
$\omega_{\mu}$ is given similarly by (see Eq.\ (21)),
\begin{equation}
     \omega_{\mu} \rightarrow \omega_{\mu} +
     \overline{\nabla}_{\mu} \phi,
\end{equation}
under which the action (28) and the two-form (30) are strict
invariants; $\omega'$ in Eq.\ (29) turns out to be invariant
modulo a total divergence. The corresponding Wilson loop for the
outer connection $\omega_{\mu}$,
\begin{equation}
     \psi'(\omega_{\mu}) = Be^{-i \int_{\Sigma} {\cal E}^{\mu\nu}
     \omega_{\mu} \ d\overline{\Sigma}_{\nu}},
\end{equation}
is also invariant under (31) in a entirely similar way to
$\psi(\rho)$ in Section VI. $\psi'$ is, of course, a topological
invariant too, and represents hence a topological phase for the
theory. \\

\noindent {\uno VII. Concluding remarks} \vspace{1em}

In this manner, we have started from topological string actions
and finished with Wilson loops as representations of the vacuum
state, that also are topological invariants in character. Hence,
as a direct consequence of such a topological invariance, the {\it
general covariance} of the theory is preserved as an unbroken
symmetry.

On the other hand, and independently on the results obtained here,
it is well known that the Wilson loops represent a possible scheme
for the quantum theories of connections such as Yang-Mills and
gravity, where all gauge invariant information in a connection is
contained in the Wilson loops. Hence, we can finish with a open
question that may be the subject of forthcoming works: may the
Wilson loops founded in the present work constitute the
fundamental blocks for a reformulation of
string theory in the search for its unbroken phase?\\

\begin{center}
{\uno ACKNOWLEDGMENTS}
\end{center}
\vspace{1em}

The author acknowledges the support from the Sistema Nacional de
Investigadores (M\'exico). \\

\end{document}